\documentclass[
showpacs,preprintnumbers,amsmath,amssymb,10pt]{revtex4}
\usepackage{amsmath}
\usepackage{graphicx}
\usepackage{subfig}
\usepackage{hyperref}
\usepackage{amssymb}
\usepackage[english]{babel}
\usepackage{epsfig}
\usepackage{wasysym}
\usepackage{graphicx}
\usepackage{indentfirst}
\usepackage{amsmath}
\usepackage{amsfonts}
\usepackage{amssymb}
\usepackage{enumerate}

\hypersetup{colorlinks,citecolor=green,filecolor=magenta,linkcolor=red,urlcolor=cyan,pdftex}

\newcommand{\be}{\begin{equation}}
\newcommand{\ee}{\end{equation}}
\newcommand{\bea}{\begin{eqnarray}}
\newcommand{\eea}{\end{eqnarray}}
\newcommand{\beaa}{\begin{eqnarray*}}
\newcommand{\eeaa}{\end{eqnarray*}}

\begin{document}
\title{Modelling{\color{white}i}of a compact{\color{white}i}anisotropic star as an{\color{white}i}anisotropic fluid{\color{white}i}sphere in $f(T)$ gravity}
\author{ D. Momeni $^{a}$, G. Abbas$^{b}$, S. Qaisar $^{c}$, Zaid Zaz$^{d}$, R. Myrzakulov $^{e}$}
\affiliation{$^{a}$ Department of Physics, College of Science,\\
Sultan Qaboos University,
P.O. Box 36, P.C. 123, Al-Khodh, Muscat, Sultanate of Oman
\\$^{b}$ Department of{\color{white}i}Mathematics,\\ The{\color{white}i}Islamia University{\color{white}i}of
Bahwalpur, Pakistan
\\$^{c}$ Department{\color{white}i}of Mathematics, COMSATS{\color{white}i}Institute of
Information{\color{white}i}Technology Sahiwal-57000, Pakistan
\\$^{d}$ Department{\color{white}i}of Electronics and{\color{white}i}Communication{\color{white}i}Engineering,
University of{\color{white}i}Kashmir, Srinagar, Kashmir-190006, India.
$^{e}$ Eurasian{\color{white}i}International Centre{\color{white}i}for Theoretical{\color{white}i}Physics and\\
Department{\color{white}i}of General \&
Theoretical{\color{white}i}Physics, Eurasian{\color{white}i}National
University, Astana 010008, Kazakhstan
}

\begin{abstract}
In{\color{white}i}this article, authors{\color{white}i}have discussed the{\color{white}i}new exact{\color{white}i}model of
anisotropic{\color{white}i}star in $f(T)$ theory of{\color{white}i}gravity. The parametric form of {\color{white}i}metric
functions has {\color{white}i}been implemented{\color{white}i}to solve{\color{white}i}the dynamical{\color{white}i}equations in $f(T)$
theory{\color{white}i}with the{\color{white}i}anisotropic fluid. The{\color{white}i}novelty of the{\color{white}i}work is{\color{white}i}that
the obtained{\color{white}i}solutions do{\color{white}i}not contain{\color{white}i}singularity but{\color{white}i}potentially
stable. The{\color{white}i}estimated values{\color{white}i}for mass and{\color{white}i}radius of the{\color{white}i}different
strange{\color{white}i}stars RX J 1856-37, Her X-1, and Vela X-12 have{\color{white}i}been
utilized{\color{white}i}to figure the{\color{white}i}values of unknown{\color{white}i}constants in Krori{\color{white}i}and
Barua{\color{white}i}metric. The physical{\color{white}i}parameters like{\color{white}i}anisotropy, stability{\color{white}i}and
redshift{\color{white}i}of the{\color{white}i}stars have been{\color{white}i}examined in{\color{white}i}detail.
\end{abstract}


\maketitle

\section{Introduction}
General{\color{white}i}Relativity (GR) is a{\color{white}i}classical gauge{\color{white}i}theory to{\color{white}i}describe long{\color{white}i}range interaction{\color{white}i}between
gravitational{\color{white}i}objects. It{\color{white}i}alone can{\color{white}i}also
account{\color{white}i}for the accelerated
expansion{\color{white}i}of the universe{\color{white}i}using a
fine- tuned{\color{white}i}cosmological constant{\color{white}i}term
$\Lambda$\cite{Sami:2009jx}-\cite{Sami:2004ic} or quintessence
scalar field{\color{white}i}as   dark{\color{white}i}energy
\cite{Geng:2015fla},\cite{Hossain:2014zma} or quintal{\color{white}i}fields
\cite{Cai:2009zp}. However{\color{white}i}the existence of{\color{white}i}dark energy{\color{white}i}requires the
introduction{\color{white}i}of additional fluids{\color{white}i}capable of
dominating{\color{white}i}over the standard{\color{white}i}pressure less{\color{white}i}matter
\cite{4}. There{\color{white}i}is a simpler{\color{white}i}course to get{\color{white}i}around dark energy{\color{white}i}and to
consistently explain{\color{white}i}the accelerated{\color{white}i}expansion. This{\color{white}i}is done{\color{white}i}by
considering a{\color{white}i}cosmological constant. However{\color{white}i}the
necessary{\color{white}i}expected value of{\color{white}i}the
cosmological constant{\color{white}i}must be extremely larger{\color{white}i}than
the observed value{\color{white}i}. This is{\color{white}i}due to quantum
considerations \cite{5}. There{\color{white}i}however is an{\color{white}i}internally consistent
approach{\color{white}i}to account for{\color{white}i}the accelerated expansion{\color{white}i}of
the universe.This is{\color{white}i}based on addressing{\color{white}i}the cosmic
dynamics by{\color{white}i}exploring further{\color{white}i}degrees of freedom{\color{white}i}of
the gravitational{\color{white}i}field. In other words{\color{white}i}to account
for the accelerated{\color{white}i}dynamics of the{\color{white}i}universe, the Einstien{\color{white}i}Hilbert
action{\color{white}i}of general relativity{\color{white}i}is
modified \cite{6}-\cite{8}. Recently a{\color{white}i}different but{\color{white}i}interesting
modification introduced as{\color{white}i}Gravity’s Rainbow{\color{white}i}models \cite{rainbow}
as{\color{white}i}an extension of{\color{white}i}the
{\color{white}i}doubly special relativity{\color{white}i}theory
\cite{double}, in which{\color{white}i}several interesting{\color{white}i}features have been
studied{\color{white}i}like absence of{\color{white}i}effective horizon in{\color{white}i}black holes,{\color{white}i}black
rings, mini black{\color{white}i}holes, gravitational{\color{white}i}collapse , brane{\color{white}i}solutions
and{\color{white}i}others \cite{faizal}.
\par
 An extensively {\color{white}i}studied approach{\color{white}i}to modified gravity{\color{white}i} stems from{\color{white}i}the "Teleparallel{\color{white}i}equivalent to General{\color{white}i}Relativity" (TEGR) \cite{Hayashi1},\cite{Hayashi2}. It is{\color{white}i}noteworthy that unlike general{\color{white}i}relativity to{\color{white}i}explain the{\color{white}i}gravitational effects{\color{white}i}TEGR uses{\color{white}i}an anti symmetric{\color{white}i}connection equipped {\color{white}i}with a{\color{white}i}non-vanishing torsion{\color{white}i}and a zero curvature (Weitzenbock connection). In other{\color{white}i}words TEGR{\color{white}i}uses torsion to{\color{white}i}explain gravitational effects{\color{white}i}unlike General{\color{white}i}relativity which uses{\color{white}i}curvature. It may{\color{white}i}be moted that {\color{white}i}in order to{\color{white}i}define the Weitzenbock connection {\color{white}i}in TEGR tetrad{\color{white}i}fields are used as{\color{white}i}dynamical variables \cite{Weitzenbock}.

Using TEGR{\color{white}i}a straight{\color{white}i}forward modification{\color{white}i}of the Einstien Hilbert{\color{white}i}action may{\color{white}i}be performed by{\color{white}i}replacing the{\color{white}i}scalar torsion{\color{white}i}with an arbitrary{\color{white}i}smoooth function{\color{white}i}of the scalar{\color{white}i}torsion $f(T)$. This{\color{white}i}class of teleparallel{\color{white}i}theories is known{\color{white}i}as $f(T)$ gravity  (see \cite{Cai:2015emx} and refrences{\color{white}i}in for a{\color{white}i}comprehensive review). Accelerated{\color{white}i}expansion of the{\color{white}i}cosmic Hubble{\color{white}i}parameter as well{\color{white}i}as dark matter{\color{white}i}can be{\color{white}i}explained using  $f(T)$ gravity \cite{Ferraro}-\cite{ Houndjo}.
\par
In gravitational{\color{white}i}theories, localized{\color{white}i}and positive{\color{white}i}valued mass
densities{\color{white}i}could suddenly collapse, emerging{\color{white}i}
thermodynamics that can{\color{white}i}produce a
{\color{white}i}hydrodynamic pressure impose{\color{white}i}on the mass
distribution{\color{white}i}. If{\color{white}i}the collapse of{\color{white}i}the
matter distribution{\color{white}i}is continued then mass{\color{white}i}contents
are transformed{\color{white}i}into a compact{\color{white}i}star. The compact{\color{white}i}star has a{\color{white}i}finite
size $R\sim 10 Km${\color{white}i}and a huge amount{\color{white}i}of
mass{\color{white}i}of order of solar mass $M\sim1.4 M_{\odot}$
\cite{book}.  A type of{\color{white}i}compact star is {\color{white}i}neutron
star{\color{white}i}, it {\color{white}i}has very{\color{white}i}strong surface
gravity and{\color{white}i}conequently a strongly coupled{\color{white}i}regime of
thermodynamics. A way to {\color{white}i}detect these{\color{white}i}massive
objects{\color{white}i} is{\color{white}i}by looking at{\color{white}i}the Doppler
shift in{\color{white}i}spectral lines emitted by{\color{white}i}atoms in the
surface{\color{white}i}of the candidated{\color{white}i}stars. One can{\color{white}i}find{\color{white}i}not
 much gravity on the{\color{white}i}earth, more on the{\color{white}i}compact
star{\color{white}i}because surface{\color{white}i}gravity of{\color{white}i}such stars is{\color{white}i}of
order $\kappa=2\times 10^{11}$. Except than{\color{white}i}this big{\color{white}i}surface
gravity,{\color{white}i}there is a{\color{white}i}greater difference between
individual plants{\color{white}i}which makes{\color{white}i}the species and{\color{white}i}neutron
stars{\color{white}i}. A reason{\color{white}i}is that the{\color{white}i}gravitational field{\color{white}i}from
compact{\color{white}i}stars, have{\color{white}i}both electric and{\color{white}i}magnetic
components{\color{white}i}and contains a{\color{white}i}definite amount{\color{white}i}of electromagnetic
energy{\color{white}i} and thanks {\color{white}i}to some{\color{white}i}basic
parameters, for{\color{white}i}example emission{\color{white}i}above-ground
level,the{\color{white}i}electromagnetic
{\color{white}i}effects{\color{white}i}are stronger
than thermodynamic{\color{white}i}effects. It was{\color{white}i}demonstrated
{\color{white}i}that the{\color{white}i}thermodynamic parameters{\color{white}i},
namely{\color{white}i}pressure $p$, energy{\color{white}i}density $\rho$, the
mass{\color{white}i}function $M$ and{\color{white}i}radius $R$ are related{\color{white}i}according to a set{\color{white}i}of
\emph{state} equations, called as Tolman-Oppenheimer-Volkoff (TOV)
equations in GR {\color{white}i}\cite{tov1}-\cite{tov3}. TOV
equation{\color{white}i}is {\color{white}i}a set of{\color{white}i}first order{\color{white}i},
non linear{\color{white}i}differential equations for a{\color{white}i}spherically
symmetric metric{\color{white}i}which is filled with{\color{white}i}the perfect
fluid with{\color{white}i}pressure and energy{\color{white}i}density. Recently{\color{white}i}TOV
equations and dynamics of{\color{white}i}neutron
{\color{white}i}stars have{\color{white}i}been investigated for{\color{white}i}different types of
modified{\color{white}i}gravity models from $f(R)$, $f(G)$ and $f(T)$ ($T$ is
torsion ) numerically and{\color{white}i}also in{\color{white}i}a non-perturbative
{\color{white}i}scheme{\color{white}i}
 \cite{Astashenok:2014pua}-
\cite{Astashenok:2014gda}.

The{\color{white}i}compact astrophysical{\color{white}i}objects such{\color{white}i}as black{\color{white}i}holes \cite{Rodrigues:2013ifa}, wormholes \cite{Bohmer:2011si},\cite{Jamil:2012ti} and compact{\color{white}i}stars have{\color{white}i}been extensively{\color{white}i}studied withtin{\color{white}i}the framework of{\color{white}i}teleparallel gravity \cite{Abbas}-\cite{Bahamonde}.
In fact{\color{white}i}compact stars{\color{white}i}have been{\color{white}i}theoritically modelled within{\color{white}i}the framework{\color{white}i}of $f(T)$ gravity. Their{\color{white}i}anisotropic behavior, regularity{\color{white}i}conditions, stability{\color{white}i}and surface{\color{white}i}redshift have been{\color{white}i}thorughly investigated. In this{\color{white}i}work we shall{\color{white}i}attempt to model{\color{white}i}anisotropic compact{\color{white}i}stars as fluid{\color{white}i}spheres within the{\color{white}i}framework of $f(T)$ gravity.  We{\color{white}i}have analysed{\color{white}i}our model for{\color{white}i}anisotropic behaviour,{\color{white}i}stablity and surface{\color{white}i}redshift.
\par
Our{\color{white}i}plan in{\color{white}i}this paper is{\color{white}i}the
following scheme:{\color{white}i}In Sec. (\ref{f(T) theory}) we
present $f(T)$ theory{\color{white}i}as an alternative{\color{white}i}theory for{\color{white}i}gravity. In Sec.
(\ref{model}) we derive equations of{\color{white}i}motion for a
spherically symmetric{\color{white}i}star. In Sec.
(\ref{analysis}) we study an{\color{white}i}isotropic model of{\color{white}i}compact star{\color{white}i}using
astrophysical data. We{\color{white}i}conclude and{\color{white}i}summarize in{\color{white}i}
Sec. (\ref{Summary and conclusion}). Some{\color{white}i}preliminary formulae{\color{white}i}are
presented in Sec. (\ref{APPENDIX}).

\section{The teleparallel equivalent of General Relativity}\label{f(T)  theory}
To{\color{white}i}begin with $f(T)$  theory,{\color{white}i}to keep away{\color{white}i}from any
perplexity,{\color{white}i}let{\color{white}i}us characterize{\color{white}i}the notion of{\color{white}i}the latin subscript $i,j$ as{\color{white}i}those
related to{\color{white}i}the tetrad{\color{white}i}field $\theta_{\mu}^{i}$ and greek  $\mu,\nu$ one identified{\color{white}i}with the space{\color{white}i}time{\color{white}i}coordinates. The line{\color{white}i}element of the{\color{white}i}manifold is{\color{white}i}given by
\begin{equation}
ds^{2}=g_{\mu\nu}dx^{\mu}dx^{\nu},
\end{equation}
The above{\color{white}i}metric can be{\color{white}i}rewritten by the{\color{white}i}tetrads
basis,defined{\color{white}i}by
\begin{equation}
dS^{2}=g_{\mu\nu}dx^{\mu}dx^{\nu}=\eta_{ij}\theta_{\mu}^{i}\theta_{\nu}^{j}dx^{\mu}dx^{\nu},
\end{equation}
\begin{equation}
dx^{\mu}=e_{i}^{\mu}\theta^{i}, \theta^{i}=e^{i}_{{\mu}}dx^{\mu},
\end{equation}
where $\eta_{ij}=diag[1,-1,-1,-1]$ and $e_{i}^{\mu}e_{i}^{\nu}=\delta_{%
\nu}^{\mu}$ or $e_{i}^{\mu}e_{j}^{\nu}=\delta_{i}^{j}$. \newline
The root of the metric determinant is given by $\sqrt{-g }%
=det[e_{\mu}^{i}]=e. $ The{\color{white}i}Weitzenb\"ock's anti-symmetric{\color{white}i}connection components{\color{white}i}in Weitzenb\"ock's{\color{white}i}spacetime \cite{Weizenbock} for{\color{white}i}vanishing
Riemann{\color{white}i}tensor part{\color{white}i}and non-vanishing{\color{white}i}torsion term{\color{white}i}are defined{\color{white}i}as
\begin{equation}
\Gamma^{\alpha}_{\mu\nu}=e_{i}^{\alpha}\partial_{\nu}e_{\mu}^{i}=-e_{i}^{%
\mu}\partial_{\nu}e_{i}^{\alpha}
\end{equation}
The{\color{white}i}torsion and{\color{white}i}the contorsion{\color{white}i}are defined{\color{white}i}by
\begin{equation}
T^{\alpha}_{\mu\nu}=\Gamma^{\alpha}_{\nu\mu}-\Gamma^{\alpha}_{\mu%
\nu}=e_{i}^{\alpha}(\partial_{\mu}e_{\nu}^{i}-\partial_{\nu}e_{\mu}^{i})
\end{equation}
\begin{equation}
K^{\mu\nu}_{\alpha}=-\frac{1}{2}(T^{\mu\nu}_{\alpha}-T^{\nu\mu}_{\alpha}-T^{%
\mu\nu}_{\alpha})
\end{equation}
and{\color{white}i}the components{\color{white}i}of the{\color{white}i}tensor $S_{\alpha}^{\mu\nu}$ as
\begin{equation}
S_{\alpha}^{\mu\nu}=\frac{1}{2}(K^{\mu\nu}_{\alpha}+\delta^{\mu}_{\alpha}T^{%
\beta\nu}_{\beta}-\delta^{\nu}_{\alpha}T^{\beta\mu}_{\beta}).
\end{equation}
Here,{\color{white}i}torsion scalar{\color{white}i}is
\begin{equation}
T=T^{\alpha}_{\mu\nu}S_{\alpha}^{\mu\nu}
\end{equation}
Analogous{\color{white}i}to $f(R)${\color{white}i}gravity,{\color{white}i}the action{\color{white}i}for $f(T)$ gravity{\color{white}i}is
\begin{equation}
S[e_{\mu}^{i}, \Phi_A]=\int d^4 x e[\frac{1}{16\pi}f(T)+\mathcal{L}%
_{\mbox{matter}}(\Phi_A)],
\end{equation}
in{\color{white}i}the above{\color{white}i}action $G=c=1$ have{\color{white}i}been used{\color{white}i}and the $\mathcal{L}%
_{matter}(\Phi_A)$ is matter{\color{white}i}field. The{\color{white}i}variation of{\color{white}i}the above{\color{white}i}action
providevthe following field{\color{white}i}equations in $f(T)${\color{white}i}gravity ,
\begin{equation}
S_{\mu}^{\nu\rho}\partial_{\rho}T
f_{TT}+[e^{-1}e^{i}_{\mu}\partial_{\rho}(ee_i^{\alpha}S_{\alpha}^{\nu%
\rho})+T^{\alpha}_{\lambda\mu} S_{\alpha}^{\nu\lambda}]f_T+\frac{1}{4}%
\delta^{\nu}_{\mu}f=4\pi \mathcal{T}_{\mu}^{\nu},
\end{equation}
where{\color{white}i}$\mathcal{T}_{\mu}^{\nu}$ is{\color{white}i}matter. In{\color{white}i}the present{\color{white}i}case, we{\color{white}i}take the
matter{\color{white}i}as anisotropic{\color{white}i}fluid for{\color{white}i}which energy-momentum{\color{white}i}tensor is
\begin{equation}
\mathcal{T}_{\mu}^{\nu}=(\rho+p_{t})u_{\mu}u^{\nu}-p_{t}\delta_{\mu}^{%
\nu}+(p_{r}-p_{t})v_{\mu}v^{\nu},
\end{equation}
where $u^{\mu}$ and $v^{\mu}$ are{\color{white}i}the four-velocity{\color{white}i}and radial-four{\color{white}i}vectors,
{\color{white}i}respectively.{\color{white}i}Further, $p_r$ and $p_t$ are{\color{white}i}pressures along{\color{white}i}radial and
tansverse{\color{white}i}directions.
\section{Model of Anisotropic Compact Stars in Generalized Telleparallel Gravity}\label{model}
We{\color{white}i}assume the{\color{white}i}geometry of{\color{white}i}star in the{\color{white}i}form of static{\color{white}i}spherically symmetric
spacetime {\color{white}i}is given{\color{white}i}
\begin{equation}
ds^{2}=e^{a(r)}dt^{2}-e^{b(r)}dr^{2}-r^{2}(d\theta ^{2}+sin^{2}(\theta
)d\phi ^{2}).  \label{3.1}
\end{equation}%
We{\color{white}i}introduce the{\color{white}i}tetrad matrix{\color{white}i}for (\ref{3.1}) as{\color{white}i}follows:
\begin{equation}
\lbrack e_{\mu }^{i}]=diag[e^{\frac{a(r)}{2}},e^{\frac{b(r)}{2}%
},r,rsin(\theta )].  \label{3.2}
\end{equation}%
One{\color{white}i}can obtain{\color{white}i}$e=det[e_{\mu }^{i}]=e^{\frac{(a+b)}{2}%
}r^{2}sin(\theta )$, and{\color{white}i}we determine{\color{white}i}the torsion{\color{white}i}scalar and
its{\color{white}i}derivative in{\color{white}i}terms of $r${\color{white}i}as
\begin{eqnarray}
T(r) &=&\frac{2e^{-b}}{r}(a^{^{\prime }}+\frac{1}{r}),  \label{3.4} \\
T^{^{\prime }}(r) &=&\frac{2e^{-b}}{r}(a^{^{\prime \prime }}+\frac{1}{r^{2}}%
-T(b^{^{\prime }}+\frac{1}{r})),
\end{eqnarray}%
where{\color{white}i}the prime{\color{white}i}denotes the{\color{white}i}derivative with{\color{white}i}respect to $r$. The{\color{white}i}set of
equations{\color{white}i}for an{\color{white}i}anisotropic fluid{\color{white}i}as \cite{Bohmer:2011si},\cite{Jamil:2012ti}
\begin{eqnarray}\label{eom1}
4\pi \rho &=&\frac{f}{4}-\frac{f_{T}}{2}\left( T-\frac{1}{r^{2}}-\frac{e^{-b}%
}{r}(a^{^{\prime }}+b^{^{\prime }})\right) ,  \label{3.8} \\ \label{eom2}
4\pi p_{r} &=&\frac{f_{T}}{2}\left( T-\frac{1}{r^{2}}\right) -\frac{f}{4}, \\ \label{eom3}
4\pi p_{t} &=&\left[ \frac{T}{2}+e^{-b}\left( \frac{a^{^{\prime \prime }}}{2}%
+(\frac{a^{^{\prime }}}{4}+\frac{1}{2r})(a^{^{\prime }}-b^{^{\prime
}})\right) \right] \frac{f_{T}}{2}-\frac{f}{4}, \\&&
\frac{cot\theta }{2r^{2}}T^{^{\prime }}f_{TT} =0\label{eom4}.
\end{eqnarray}%

Further,{\color{white}i}the last{\color{white}i}equation leads{\color{white}i}to the the{\color{white}i}following linear{\color{white}i}form of $f(T)$ :
\begin{equation}
f(T)=\beta T+{\beta }_{1},  \label{3.9}
\end{equation}%
where $\beta $ and ${\beta }_{1}$ are{\color{white}i}integration constants.{\color{white}i}We parameterize
the{\color{white}i}metric as the{\color{white}i}following:
\begin{eqnarray}
e^{a(r)}=( 1+br^{2}) ^{2},e^{b(r)}=\left( A+\frac{B(
2+br^{2}) ^{3/2}}{3\sqrt{b}}\right) ^{2}
\end{eqnarray}%
where $b$ is{\color{white}i}a {\color{white}i}constant, the{\color{white}i}arbitrary constants $A$, $B$ and $b$ can be{\color{white}i}evaluated by{\color{white}i}using some
physical{\color{white}i}matching conditions. Now{\color{white}i}using equations{\color{white}i}given in (\ref{eom1}-\ref{eom4}), we{\color{white}i}will  get
forms{\color{white}i}of matter components{\color{white}i}given in{\color{white}i}equations  (\ref{ap1}-\ref{ap7}) in appendix Sec. (\ref{APPENDIX}).

\begin{figure}[tbp]
\center\epsfig{file=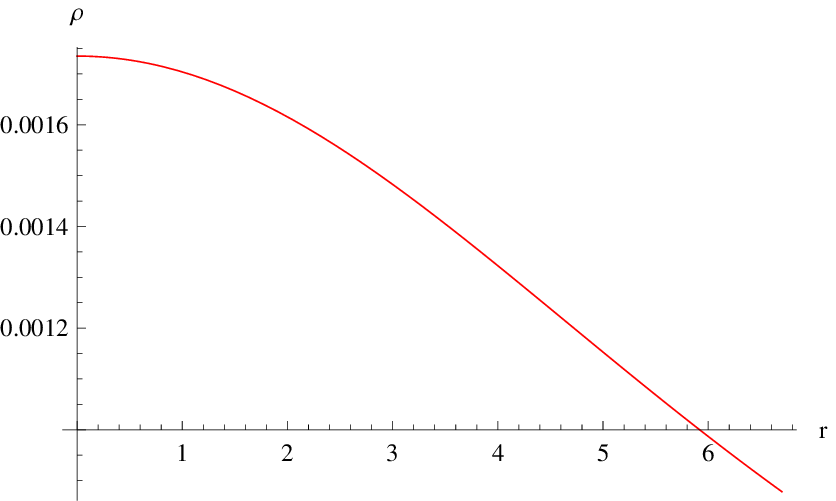, width=0.3\linewidth} \epsfig{file=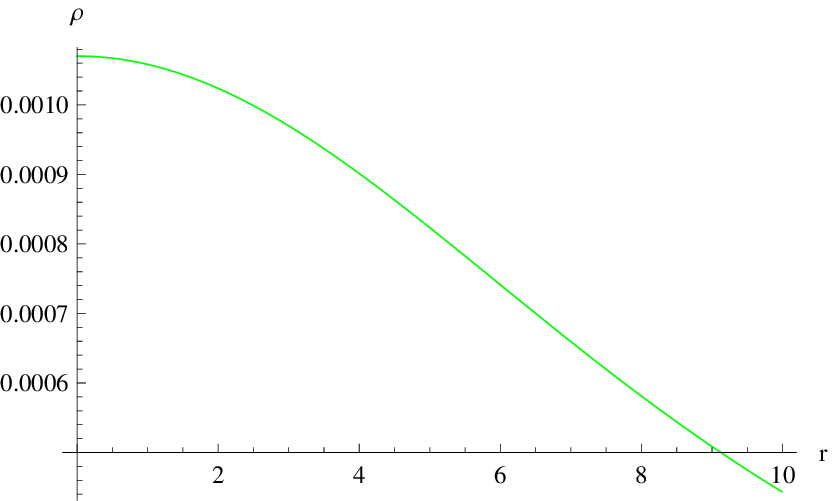,
width=0.3\linewidth} \epsfig{file=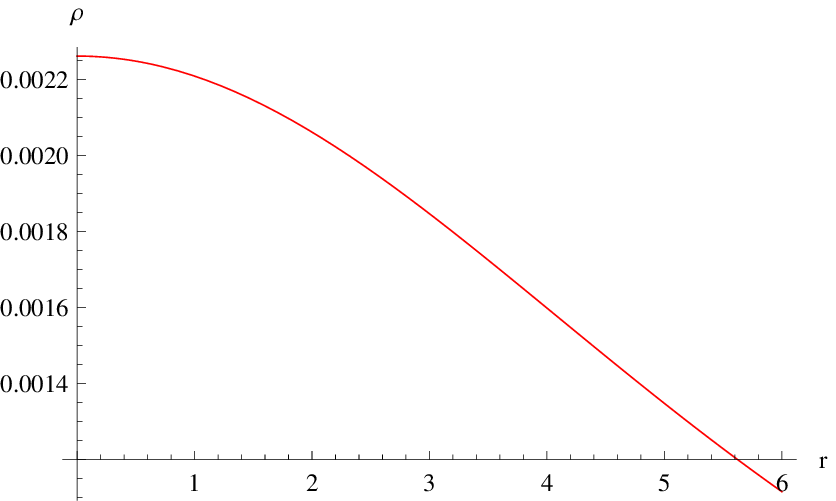, width=0.3\linewidth}
\caption{ Density variation for Strange star candidate RX J 1856-37,
Her X-1, and Vela X-12, respectively.}
\end{figure}

\begin{figure}[tbp]
\center\epsfig{file=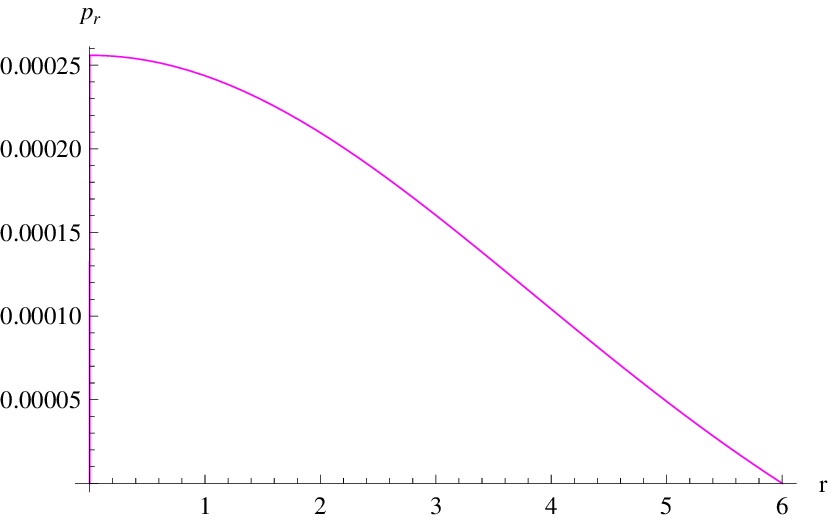, width=0.3\linewidth} \epsfig{file=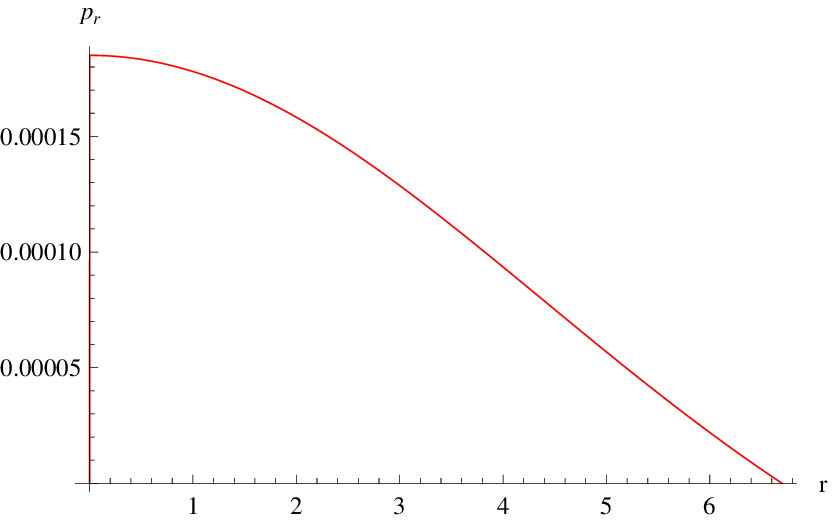,
width=0.3\linewidth} \epsfig{file=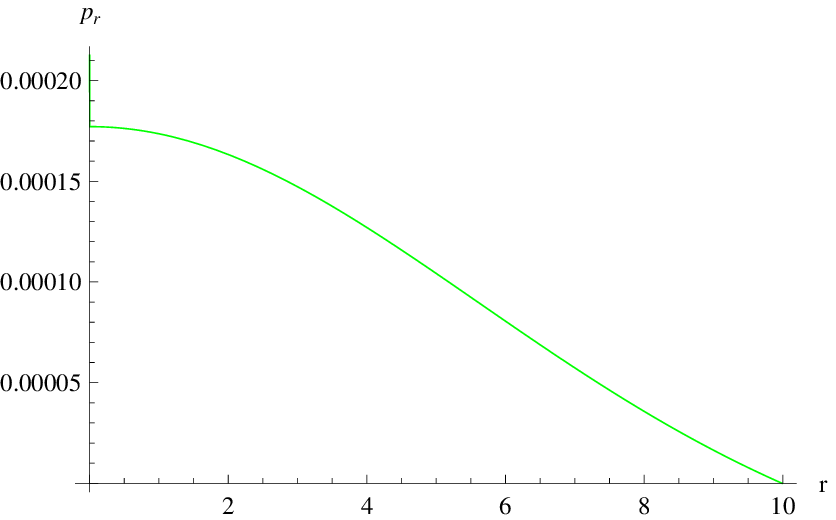, width=0.3\linewidth}
\caption{Radial pressure variation for Strange star candidate RX J
1856-37, Her X-1, and Vela X-12, respectively.}
\end{figure}

\begin{figure}[tbp]
\center\epsfig{file=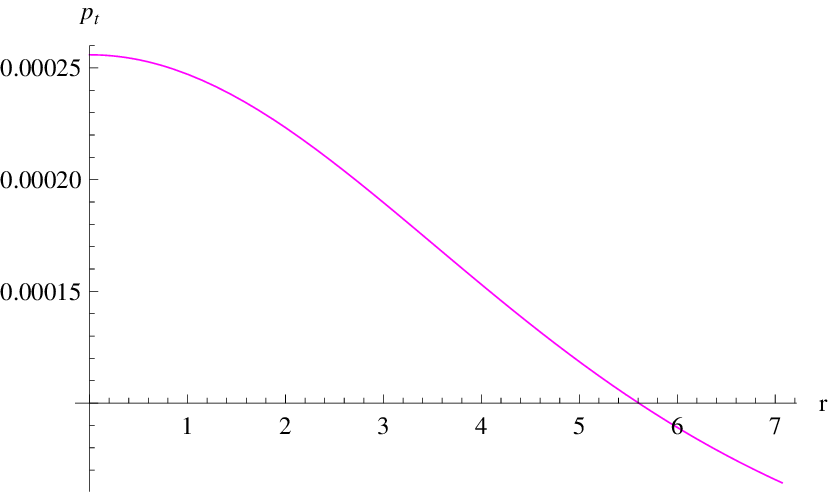, width=0.3\linewidth} \epsfig{file=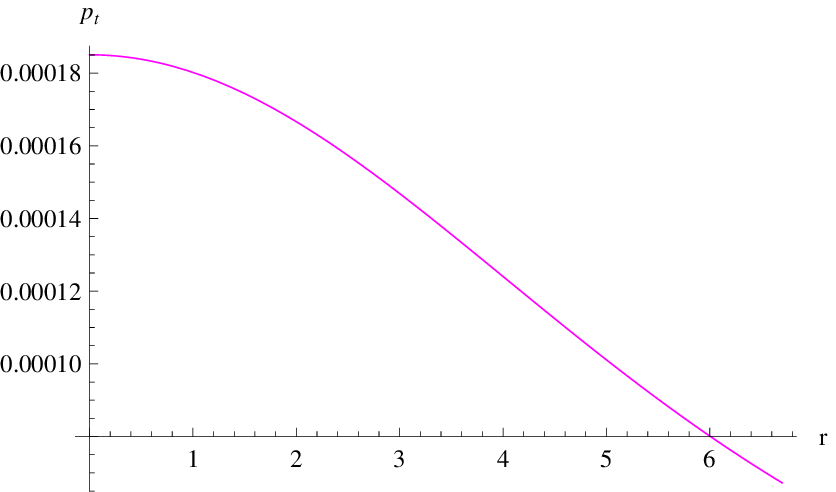,
width=0.3\linewidth} \epsfig{file=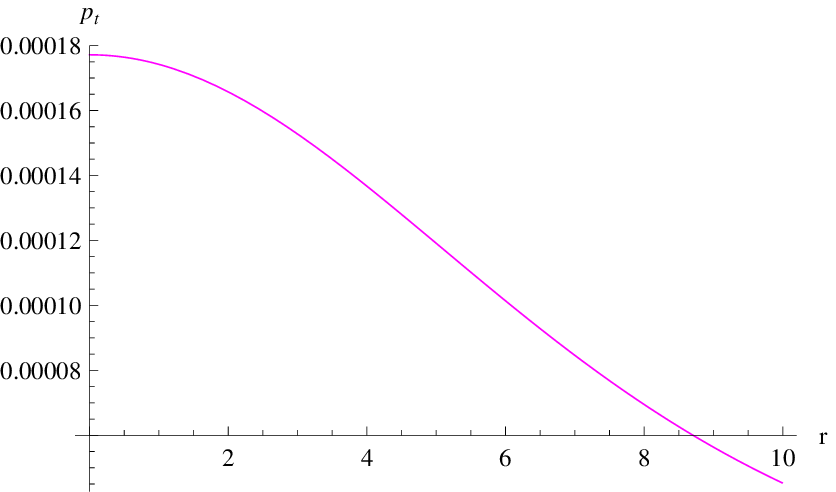, width=0.3\linewidth}
\caption{ Transverse pressure variation for Strange star candidate
RX J 1856-37, Her X-1, and Vela X-12, respectively.}
\end{figure}

\begin{figure}[tbp]
\caption{ EOS parameter ${%
\protect\omega }_{r}$ variation for Strange star candidate RX J
1856-37, Her
X-1, and Vela X-12 , respectively.}\center%
\epsfig{file=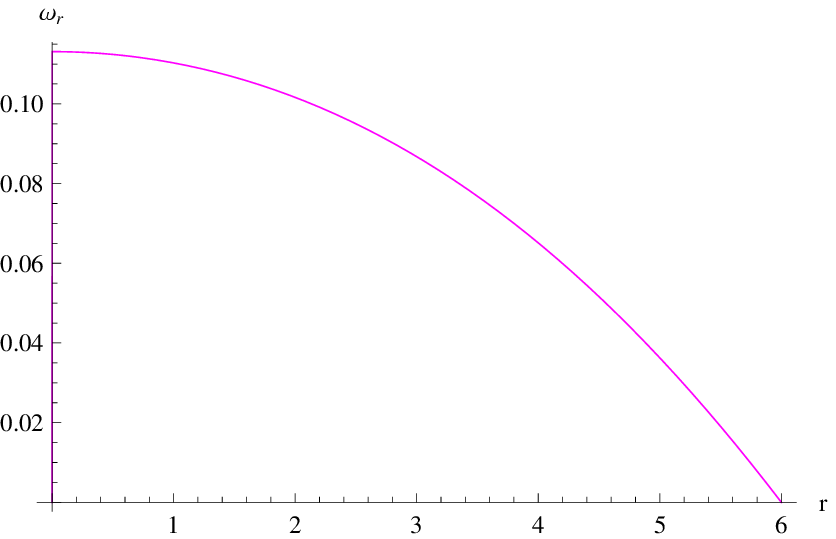,
width=0.3\linewidth} \epsfig{file=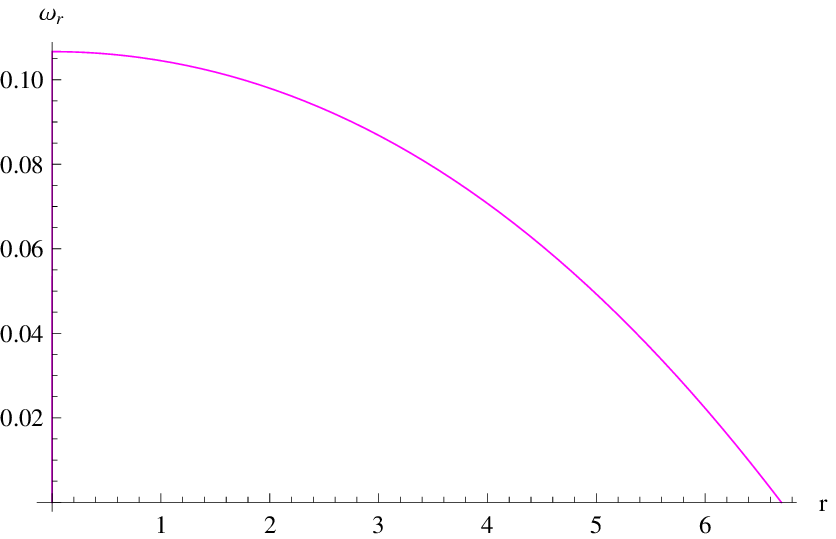,
width=0.3\linewidth} \epsfig{file=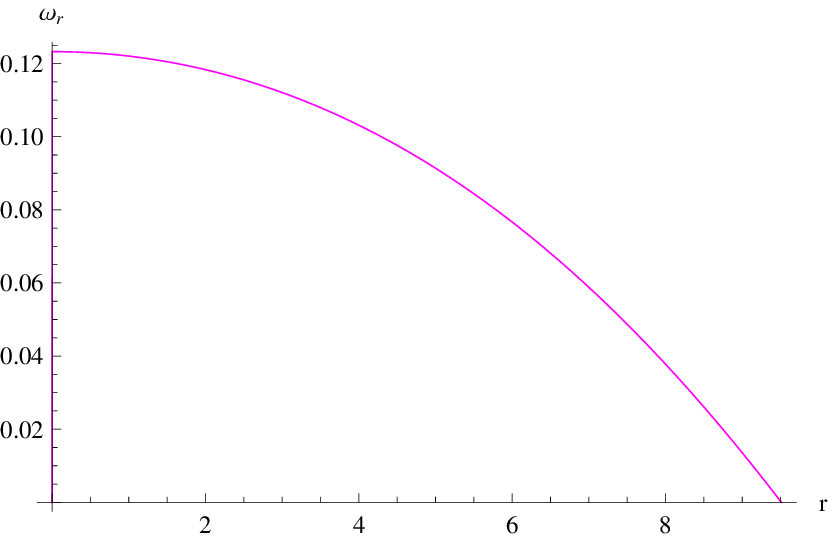,
width=0.3\linewidth}\center\epsfig{file=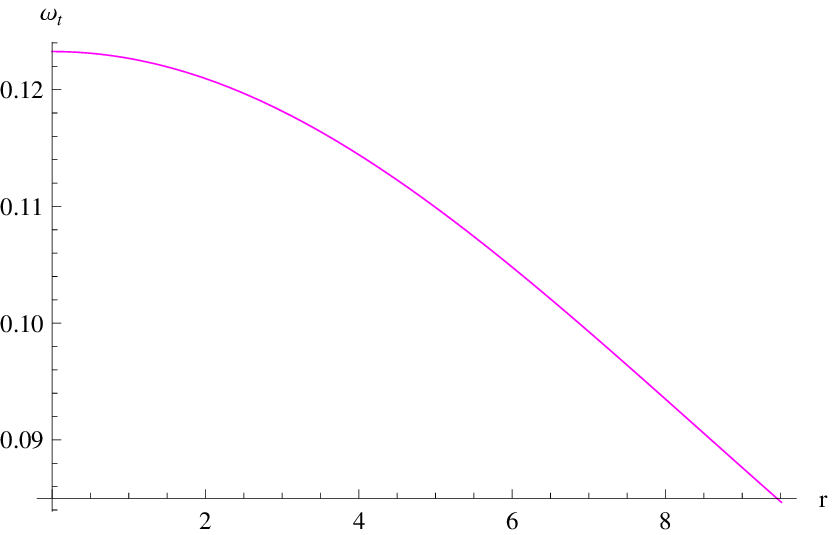, width=0.3\linewidth} %
\epsfig{file=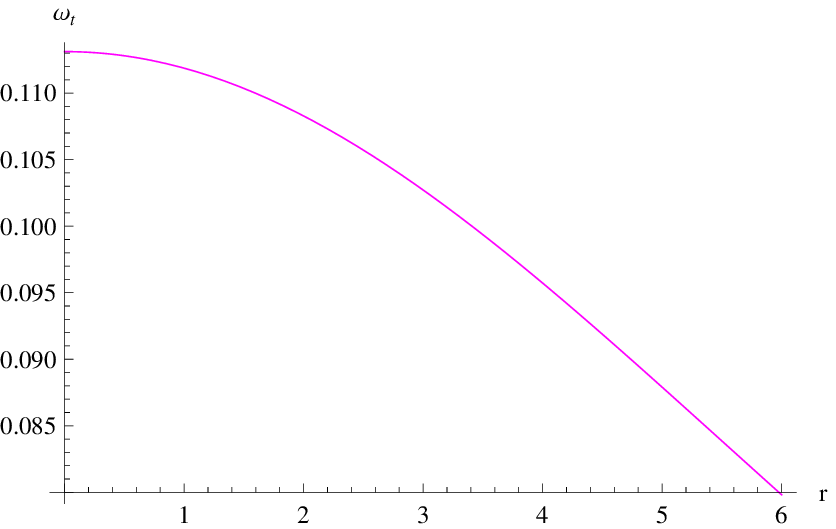, width=0.3\linewidth}
\epsfig{file=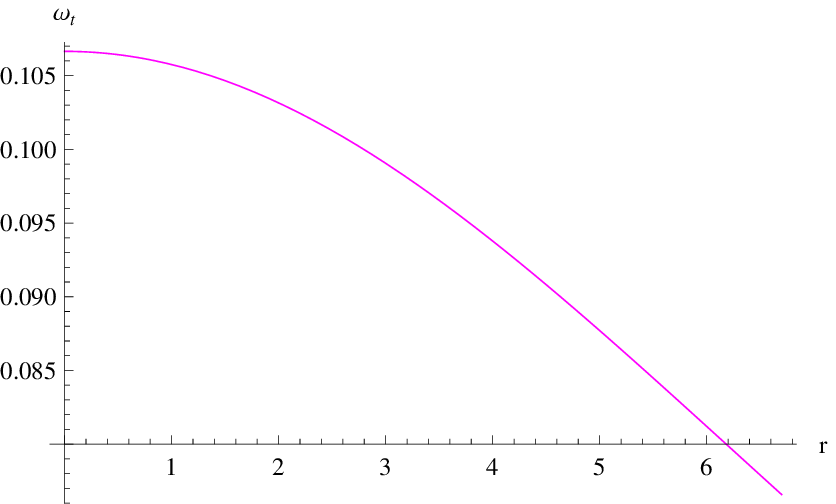,
width=0.3\linewidth}
\caption{ EOS parameter ${%
\protect\omega }_{t}$ variation of Strange star candidate Cen-X-3, RX J
1856-37 and Her X-1, respectively.}
\end{figure}

The{\color{white}i}behavior of{\color{white}i}density, radial{\color{white}i}and transverse{\color{white}i}pressures and
equation{\color{white}i}of state (EoS){\color{white}i}parameters are{\color{white}i}given in{\color{white}i}figures
\textbf{1-5} for
a strange{\color{white}i}star candidate{\color{white}i}RX J 1856-37,{\color{white}i}Her X-1, and{\color{white}i}Vela X-12,{\color{white}i}respectively.{\color{white}i}In this{\color{white}i}case EoS{\color{white}i}parameters have{\color{white}i}values
${\omega}_t>0$ and $0<{\omega}_r \leq1$, which{\color{white}i}shows the{\color{white}i}fact that
star{\color{white}i}consists of{\color{white}i}ordinary matter{\color{white}i}and effect of $f(T)$ model{\color{white}i}in the
present{\color{white}i}setup.

\newpage
\section{ Analysis of the Proposed Model}\label{analysis}

Here, we{\color{white}i}discuss the following{\color{white}i}properties of{\color{white}i}the proposed{\color{white}i}model:

\subsection{Anisotropic Behavior}

From{\color{white}i}Eqs.(\ref{ap1}-\ref{ap7}), we{\color{white}i}get the radial{\color{white}i}gradient of{\color{white}i}pressures . We{\color{white}i}compute second{\color{white}i}order derivatives  $\frac{d^{2}\rho }{dr^{2}}$. We{\color{white}i}observe that{\color{white}i}at center $r=0$, our{\color{white}i}model provides{\color{white}i}that
\begin{eqnarray}  \label{se4}
\frac{d\rho}{dr}&=&0,~~~ \frac{dp_r}{dr}=0 \\
\frac{d^2\rho}{dr^2}&<&0,~~~~ \frac{d^2p_r}{dr^2}<0.
\end{eqnarray}
This{\color{white}i}indicate maximality{\color{white}i}of radial{\color{white}i}pressure and{\color{white}i}density. This{\color{white}i}fact
implies{\color{white}i}that $\rho$ and $p_r$ are{\color{white}i}decreasing function{\color{white}i}of $r$ as
shown in{\color{white}i}figures \textbf{1,2}{\color{white}i}for a{\color{white}i}class of strange{\color{white}i}stars.

\begin{figure}[tbp]
\center\epsfig{file=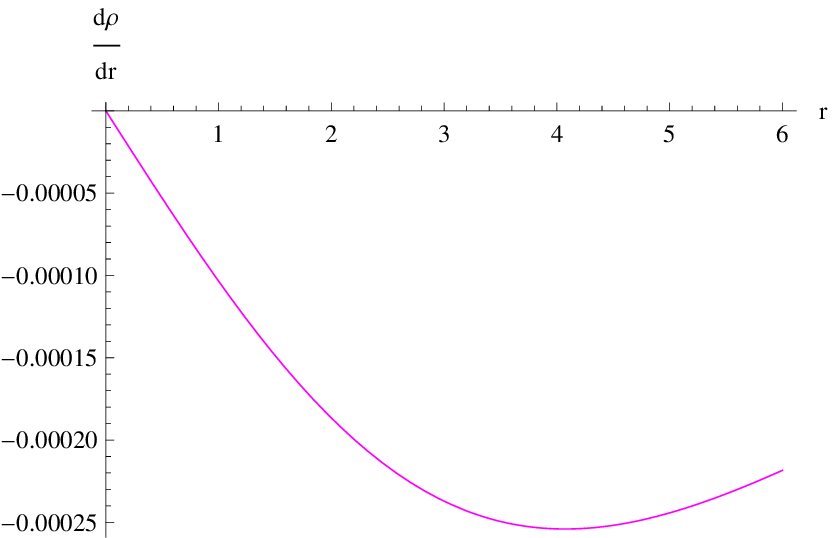, width=0.45\linewidth}
\epsfig{file=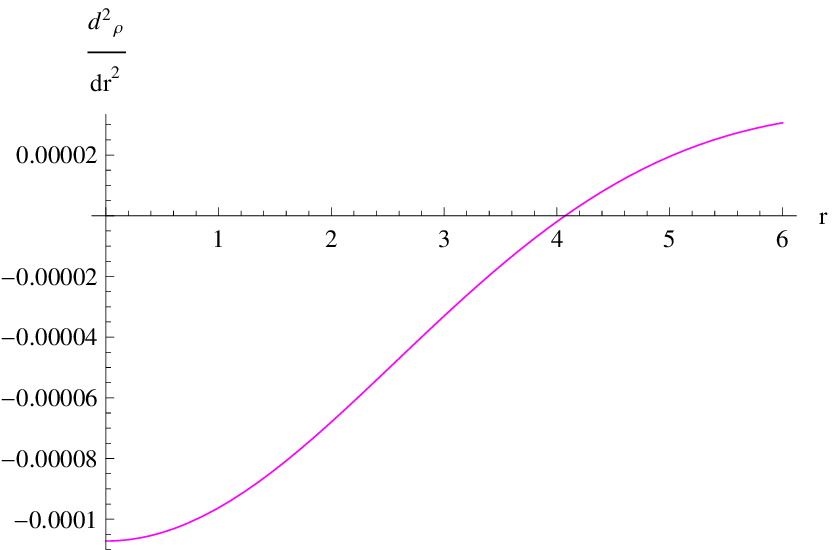, width=0.45\linewidth} \epsfig{file=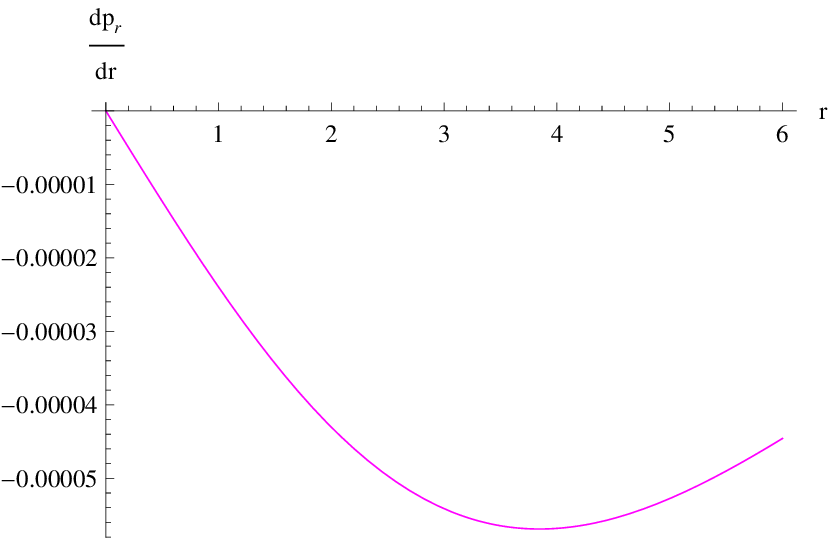,
width=0.45\linewidth}\epsfig{file=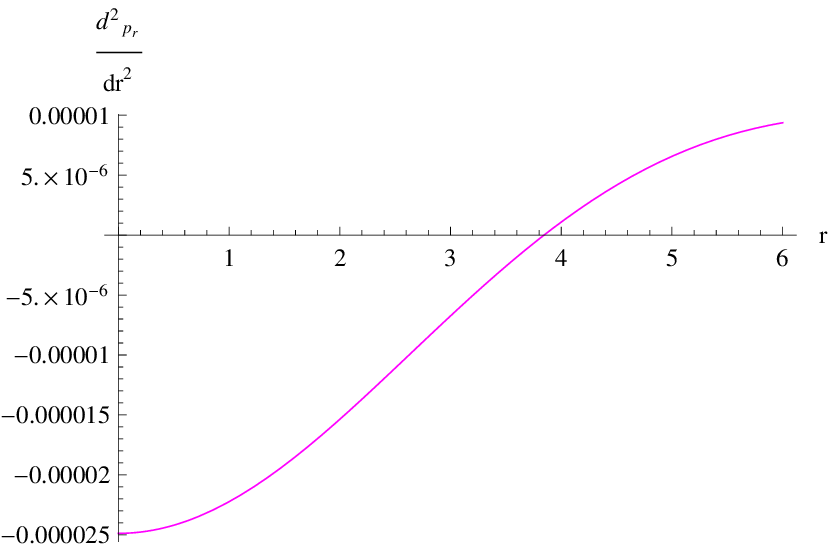, width=0.45\linewidth}
\caption{The plotted graphs are only for the data of 4U 1820-30.}
\end{figure}

\begin{figure}[tbp]
\center\epsfig{file=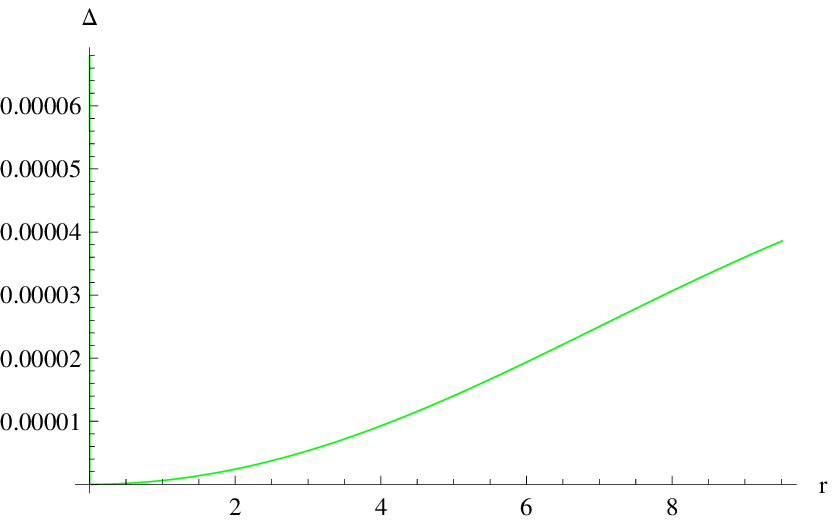, width=0.3\linewidth}
\epsfig{file=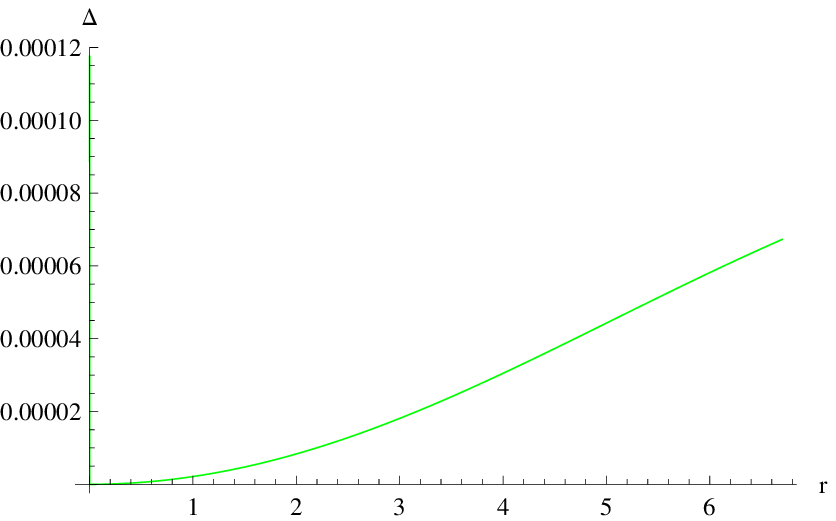, width=0.3\linewidth} \epsfig{file=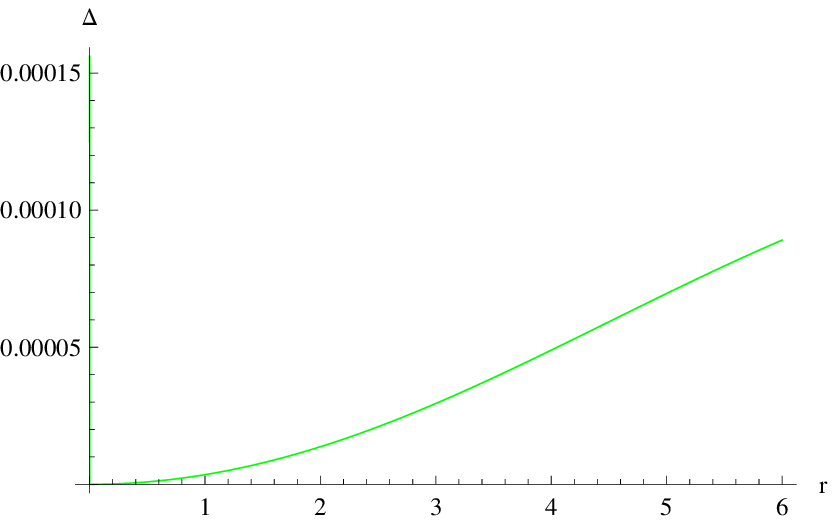,
width=0.3\linewidth} \caption{ Variation of anisotropy $\Delta$ for
Strange star candidate Her X-1, SAX J 1808.4-3658(SS1) and 4U
1820-30, respectively.}
\end{figure}
The
measure{\color{white}i}of anisotropy{\color{white}i}is
\begin{equation*}
\Delta=\frac{2}{r}(p_t-p_r),
\end{equation*}
which{\color{white}i}takes the form{\color{white}i}given in Eq. (\ref{ap10}).

It is{\color{white}i}well known{\color{white}i}that anisotropy{\color{white}i}will be directed{\color{white}i}outward when $p_{t}>p_{r}$
i.e., $\Delta >0$, and inward when $p_{t}<p_{r}$ i.e., $\Delta <0$. It is
apparent{\color{white}i}from the{\color{white}i}figure \textbf{7} that{\color{white}i}for our{\color{white}i}model that a{\color{white}i}repulsive
(anisotropic) force would exists as $(\Delta >0)$ (for smaller values of $r$%
) which{\color{white}i}permits the{\color{white}i}formation of super{\color{white}i}massive star, while{\color{white}i}for larger{\color{white}i}values
of $r$, $\Delta =0$, where{\color{white}i}a star comes{\color{white}i}to the equilibrium{\color{white}i}position.

The{\color{white}i}anisotropy will{\color{white}i}be directed{\color{white}i}outward when $p_t>p_r$ this implies{\color{white}i}that $%
\Delta>0$ and{\color{white}i}directed inward{\color{white}i}when $p_t<p_r$ implying{\color{white}i}$\Delta>0$. In
this{\color{white}i}case $\Delta>0$, for{\color{white}i}larger values{\color{white}i}of $r$ for a{\color{white}i}class of
strange{\color{white}i}stars as{\color{white}i}shown in{\color{white}i}figures \textbf{7}. This{\color{white}i}implies that
anisotropic{\color{white}i}force allows{\color{white}i}the construction{\color{white}i}of more{\color{white}i}massive stars{\color{white}i}.
\par
 In
order{\color{white}i}to comprehend{\color{white}i}some general{\color{white}i}results associated{\color{white}i}with the{\color{white}i}strong
gravitational{\color{white}i}fields, we{\color{white}i}include weakvenergy condition{\color{white}i}(WEC), null
energy{\color{white}i}condition (NEC), strong{\color{white}i}energy condition{\color{white}i}(SEC) and{\color{white}i}dominant
energy{\color{white}i}condition (DEC).{\color{white}i}For an anisotropic{\color{white}i}fluid, these{\color{white}i}are defined
as
\begin{eqnarray}
\mathbf{NEC}:\quad&&\rho+p_r\geq0, \quad \rho+p_t\geq0,  \notag \\
\mathbf{WEC}:\quad&&\rho\geq0, \quad \rho+p_r\geq0, \quad \rho+p_t\geq0,
\notag \\
\mathbf{SEC}:\quad&&\rho+p_r\geq0, \quad \rho+p_t\geq0, \quad
\rho+p_r+2p_t\geq0,  \notag \\
\mathbf{DEC}:\quad&&\rho>|p_r|, \quad \rho>|p_t|.  \notag
\end{eqnarray}
We find{\color{white}i}that our model{\color{white}i}satisfies these{\color{white}i}conditions .

\subsection{Matching{\color{white}i}of interior{\color{white}i}and exteriorvspacetime}

The intrinsic{\color{white}i}metric of the{\color{white}i}boundary surface will{\color{white}i}be the same
whether it is derived{\color{white}i}from the internal or{\color{white}i}external geometry of the
star. It{\color{white}i}proves that for any{\color{white}i}well defined coordinate{\color{white}i}system, the metric{\color{white}i}tensor
components will be{\color{white}i}continuous across the boundary{\color{white}i}surface. for this
purpose the{\color{white}i}matching conditions are equested for the interior metric
. In this case, one{\color{white}i}can solve the{\color{white}i}system of field{\color{white}i}equations with{\color{white}i}auxiliary
 boundary condition $p_r(r=R) =
0$. It is needed to match{\color{white}i}smoothly the  interior{\color{white}i}metric  to an{\color{white}i}exterior Schwarzschild{\color{white}i}metric given by
\begin{equation}  \label{21}
ds^2=\left(1-\frac{2M}{r}\right)dt^2-{\color{white}i}\left(1- \frac{2M}{r}%
\right)^{-1}dr^2-r^2(d\theta^2+\sin^2\theta d\varphi^2).
\end{equation}
At the surface{\color{white}i}of star,  $r=R$ continuity of the metric{\color{white}i}functions $g_{tt}$, $g_{rr}$
and $\frac{\partial g_{tt}}{\partial r}$ at the boundary{\color{white}i}surface yield,
\begin{eqnarray}  \label{22}
g_{tt}^-=g_{tt}^+,~~~~~ g_{rr}^-=g_{rr}^+,~~~~~ \frac{\partial g_{tt}^-}{%
\partial r}=\frac{\partial g_{tt}^+}{\partial r},
\end{eqnarray}
where $-$ and $+$, correspond{\color{white}i}to interior{\color{white}i}and exterior{\color{white}i}solutions.
By matching the interior solution $\left( \ref{3.1}\right) $ and exterior
solution $\left( \ref{4.6}\right) $ at the boundary{\color{white}i}surface $r=r_{c}$ ,
we obtain following{\color{white}i}three equations
\begin{eqnarray}
e^{a_c} &=&1-\frac{2M}{r_{c}}=\left( A+\frac{B\left( 2+br^{2}\right) ^{3/2}}{3%
\sqrt{b}}\right) ^{2},  \label{4.8} \\
e^{-b_c} &=&1-\frac{2M}{r_{c}}=\left( 1+br_{c}^{2}\right) ^{-2}  \label{4.9}
\\
p_{r}\left( r_{c}\right) &=&0  \label{4.10}
\end{eqnarray}

\bigskip Using the{\color{white}i}boundary condition $\left( \ref{4.8}-\ref{4.10}\right) $
we obtain

\begin{equation}
\frac{A}{B}=\frac{2-4r_{c}^{2}-b^{2}r_{c}^{4}}{3\sqrt{b}\sqrt{2+br_{c}^{2}}}
\label{4.11}
\end{equation}

\begin{equation}
A=\left( 1+br_{c}^{2}\right) ^{-1}-\frac{B\left( 2+br_{c}^{2}\right) ^{3/2}}{%
3\sqrt{b}}  \label{4.12}
\end{equation}

\begin{equation}
\frac{2M}{r_{c}}=1-\left( 1+br_{c}^{2}\right) ^{-2}  \label{4.13}
\end{equation}

Solving $\left( \ref{4.11}-\ref{4.13}\right) $ we obtain,
\begin{equation}
B=\frac{\sqrt{b}}{2}\frac{\sqrt{2+br_{c}^{2}}}{1+br_{c}^{2}}
\end{equation}

\begin{equation}
A=\frac{2-2-4r_{c}^{2}-b^{2}r_{c}^{4}}{6\left( 1+br_{c}^{2}\right) }
\end{equation}

\begin{equation}
b=\frac{1}{r_{c}^{2}}\left[ \frac{1}{\sqrt{1-\frac{2M}{r_{c}}}}-1\right]
\end{equation}

Now{\color{white}i}the values{\color{white}i}of constants $A,$ $b$ and $B$ for some well known{\color{white}i}compact
stars are{\color{white}i}obtained in{\color{white}i}table \textbf{1}.

\begin{table}[th]
\caption{The values of constants are{\color{white}i}obtained from model for some
well known model compact stars}
\begin{center}
\begin{tabular}{|c|c|c|c|c|c|c|}
\hline
{Compact Star} & \textbf{\ $Mass$} & \textbf{$Radius(km)$} & \textbf{\ }$%
b(km^{-2})$ & \textbf{\ $B(km^{-1})$} & $A$ & $\frac{M}{r}$ \\ \hline
Her X-1 & 0.98$M_{\odot }$ & 6.7 & 0.007268 & 0.049020 & $0.073950$ &
0.215746 \\ \hline
RX J 1856-37 & 0.9031$M_{\odot }$ & 6 & 0.009475 & 0.055529 & $0.064509$ &
0.222012 \\ \hline
Vela X-12 & 1.77$M_{\odot }$ & 9.99 & 0.004483 & 0.036184 & $0.001172$ &
0.261336 \\ \hline
Cen X-3 & 1.49$M_{\odot }$ & 9.51 & 0.004020 & 0.035744 & $0.050524$ &
0.231098 \\ \hline
\end{tabular}%
\end{center}
\end{table}

For{\color{white}i}the given{\color{white}i}values of $M$ and $R$ for given{\color{white}i}star, the{\color{white}i}constants
$A$ and $B$ are{\color{white}i}given in the{\color{white}i}table \textbf{1}. We{\color{white}i}would like to
mention{\color{white}i}that the values{\color{white}i}of $M$ and $R$ have{\color{white}i}been taken{\color{white}i}from \cite{Lattimer
et al. (2014)}, \cite{Li et al. (1999)}, \cite{Leahy et al. (2007)} to calculate
the values of $A$ and $B$.

\subsection{Stability}

Here,{\color{white}i}we discuss the{\color{white}i}stability of{\color{white}i}strange stars{\color{white}i}for this purpose{\color{white}i}the  radial{\color{white}i}sound
speed $\upsilon _{sr}$ and{\color{white}i}transverse sound{\color{white}i}speed $\upsilon _{st}^{2} $  are given by

\begin{equation*}
\upsilon _{sr}^{2}=\frac{dp_{r}}{d\rho }\equiv \frac{-\left( 3\left(
\begin{array}{c}
108A^{3}b^{3/2}\left( 2+br^{2}\right) ^{3/2}+9A^{2}bB\left( 1+br^{2}\right)
\left( 56+br^{2}\left( 47+9br^{2}\right) \right) \\
-2B^{3}\left( 2+br^{2}\right) ^{3}\left( 10+br^{2}\left( 7+br^{2}\left(
2+br^{2}\right) \right) \right) \\
+3A\sqrt{b}B^{2}\left( 2+br^{2}\right) ^{3/2}\left( 4+br^{2}\left(
41+br^{2}\left( 28+3br^{2}\right) \right) \right)%
\end{array}%
\right) \right) }{%
\begin{array}{c}
2\left( 2+br^{2}\right) \left( 3A\sqrt{b}+B\left( 2+br^{2}\right)
^{3/2}\right) \\
\times \left(
\begin{array}{c}
9A^{2}b\sqrt{2+br^{2}}\left( 12+\left( 1+br^{2}\right) \left(
3+br^{2}\right) \beta \right) \\
+B^{2}\left( 2+br^{2}\right) ^{3/2} \\
\left( 78-18\beta +br^{2}\left(
\begin{array}{c}
96-20\beta + \\
br^{2}\left( 30+\left( 5+br^{2}\left( 8+br^{2}\right) \right) \beta \right)%
\end{array}%
\right) \right) \\
+3A\sqrt{b}B\left( 3(39+\beta )+br^{2}\left(
\begin{array}{c}
126+26\beta \\
+br^{2}\left( 33+\left( 37+2br^{2}\left( 8+br^{2}\right) \right) \beta
\right)%
\end{array}%
\right) \right)%
\end{array}%
\right)%
\end{array}%
}
\end{equation*}

\begin{eqnarray*}
\upsilon _{st}^{2} &=&\frac{dp_{t}}{d\rho } \\
&\equiv &-\frac{\left( 1+br^{2}\right) \left( 9A^{2}b\sqrt{2+br^{2}}\left(
3+br^{2}\right) +B^{2}\left( 2+br^{2}\right) ^{3/2}\left( -18+br^{2}\left(
-2+br^{2}\left( 7+br^{2}\right) \right) \right) \right) \beta }{\left(
\begin{array}{c}
9A^{2}b\sqrt{2+br^{2}}\left( 12+\left( 1+br^{2}\right) \left(
3+br^{2}\right) \beta \right) \\
+B^{2}\left( 2+br^{2}\right) ^{3/2}\left( 78-18\beta +br^{2}\left(
96-20\beta +br^{2}\left( 30+\left( 5+br^{2}\left( 8+br^{2}\right) \right)
\beta \right) \right) \right) \\
+3A\sqrt{b}B\left( 3(39+\beta )+br^{2}\left( 126+26\beta +br^{2}\left(
33+\left( 37+2br^{2}\left( 8+br^{2}\right) \right) \beta \right) \right)
\right)%
\end{array}%
\right) }
\end{eqnarray*}

\begin{equation*}
-\frac{\left( 1+br^{2}\right) \left( 3A\sqrt{b}B\left( 3+br^{2}\left(
23+2br^{2}\left( 7+br^{2}\right) \right) \right) \right) \beta }{\left(
\begin{array}{c}
9A^{2}b\sqrt{2+br^{2}}\left( 12+\left( 1+br^{2}\right) \left(
3+br^{2}\right) \beta \right) \\
+B^{2}\left( 2+br^{2}\right) ^{3/2}\left( 78-18\beta +br^{2}\left(
96-20\beta +br^{2}\left( 30+\left( 5+br^{2}\left( 8+br^{2}\right) \right)
\beta \right) \right) \right) \\
+3A\sqrt{b}B\left( 3(39+\beta )+br^{2}\left( 126+26\beta +br^{2}\left(
33+\left( 37+2br^{2}\left( 8+br^{2}\right) \right) \beta \right) \right)
\right)%
\end{array}%
\right) }
\end{equation*}


Reference{\color{white}i}\cite{Herrera (1992)} developed a {\color{white}i}proposal to{\color{white}i}check the
stability of anisotropic gravitating{\color{white}i}source. Currently, this{\color{white}i}technique is
termed as cracking concept{\color{white}i}which states{\color{white}i}that if radial speed of{\color{white}i}sound is
greater than{\color{white}i}the transverse{\color{white}i}speed of{\color{white}i}sound in a region then{\color{white}i}such a region is
a potentially{\color{white}i}stable region,{\color{white}i}otherwise unstable{\color{white}i}region. In our case, figure
\textbf{8} indicates that there{\color{white}i}is change{\color{white}i}of sign for{\color{white}i}the term $\upsilon
_{st}^{2}-\upsilon _{sr}^{2}$ within the specific{\color{white}i}configuration.  We{\color{white}i}can see that $\mid\upsilon^2_{st}-%
\upsilon^2_{sr}\mid\leq1$. This is{\color{white}i}used to check{\color{white}i}whether local
anisotropic matter{\color{white}i}distribution is{\color{white}i}stable or not. Hence, we
conclude that our{\color{white}i}strange star{\color{white}i}model is{\color{white}i}unstable.

\begin{figure}[tbp]
\center\epsfig{file=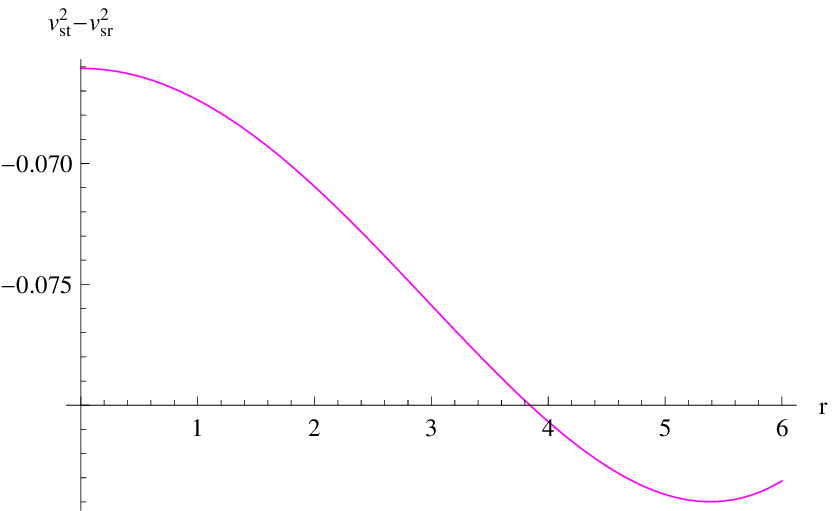, width=0.3\linewidth}
\epsfig{file=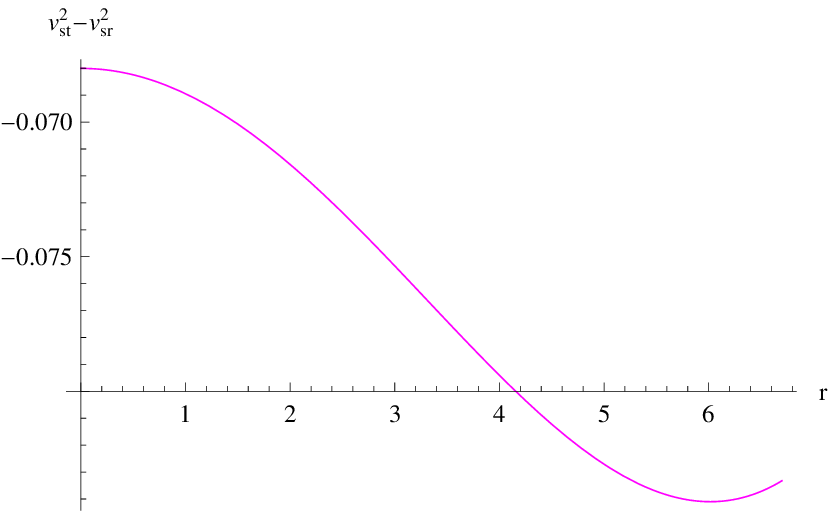,
width=0.3\linewidth} \epsfig{file=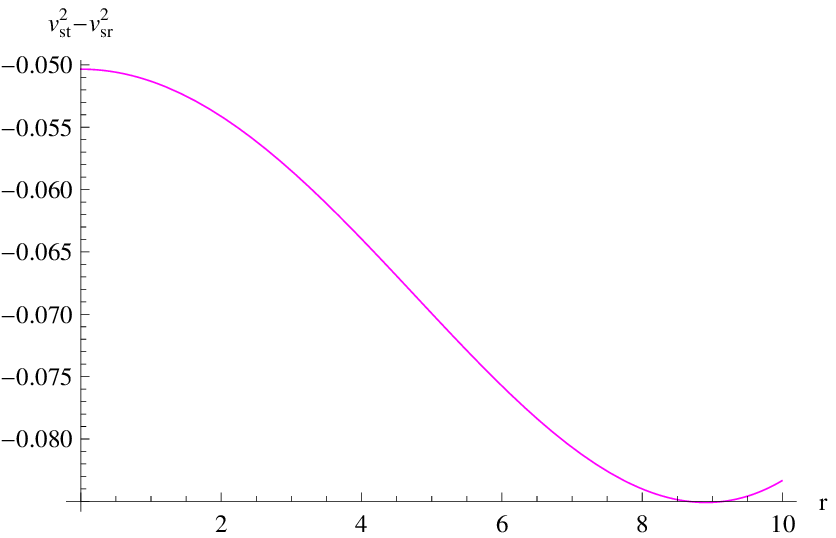,
width=0.3\linewidth}
\caption{ Variation of $\protect%
\upsilon _{st}^{2}-\protect\upsilon _{sr}^{2}$ for Strange star candidate
Her X-1, SAX J 1808.4-3658(SS1) and 4U 1820-30, respectively.}
\end{figure}

\begin{figure}[tbp]
\center\epsfig{file=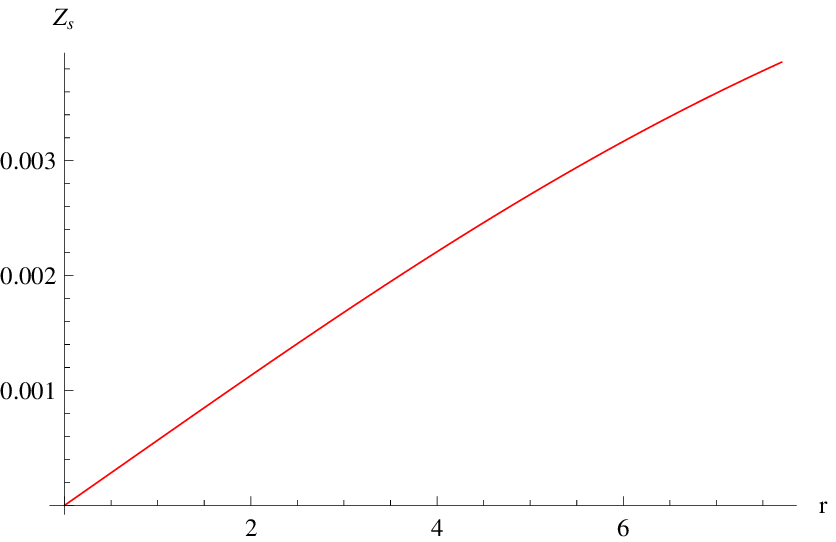, width=0.3\linewidth}
\epsfig{file=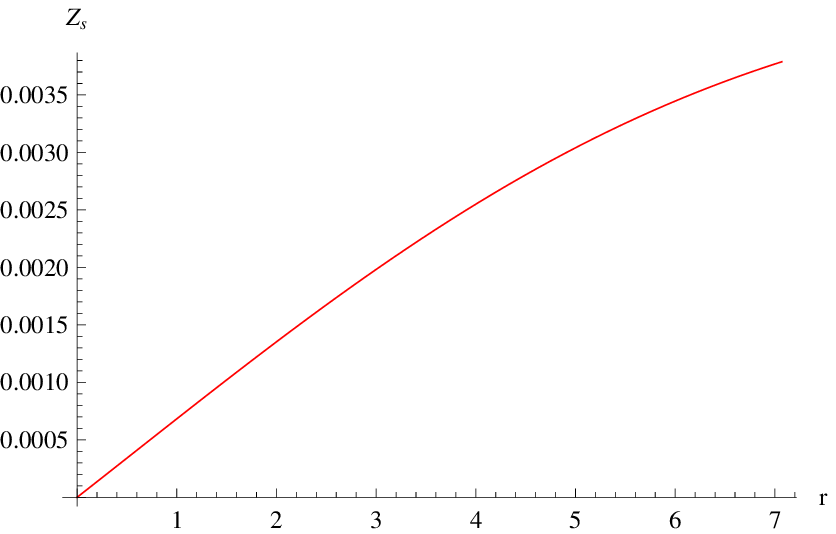, width=0.3\linewidth} \epsfig{file=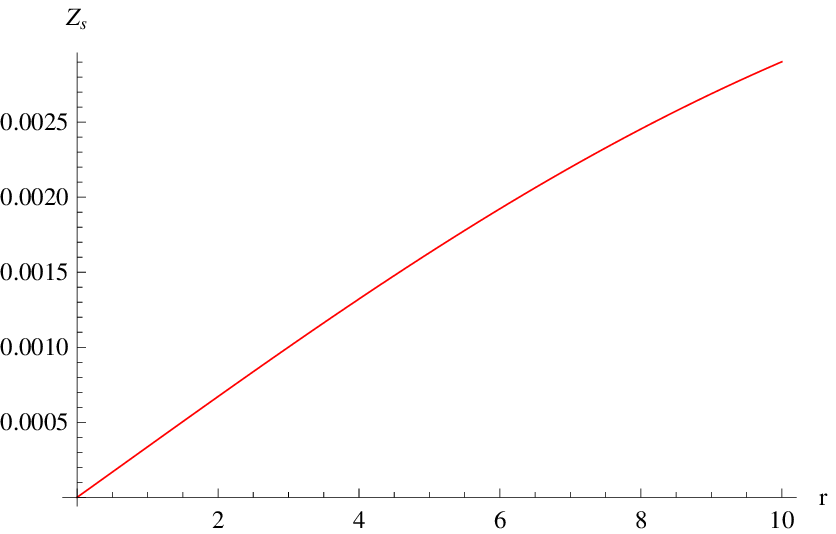,
width=0.3\linewidth} \caption{ Variation of surface redshift $Z_s$
for Strange star candidate Her X-1, SAX J 1808.4-3658(SS1) and 4U
1820-30, respectively.}
\end{figure}
\subsection{Surface Redshift}

The{\color{white}i}compactness of{\color{white}i}the star is{\color{white}i}given by
\begin{equation}
u\equiv e^{-a}=1-\frac{2m}{r}
\end{equation}

In our{\color{white}i}model the{\color{white}i}effective mass is{\color{white}i}given by

\begin{equation*}
m\left( r\right) =4\pi \int_{0}^{r}\rho r^{2}dr
\end{equation*}

The{\color{white}i}surface redshift $(Z_{s})$ resulting{\color{white}i}from the{\color{white}i}compactness $u$ is
obtained as
\begin{equation}
1+Z_{s}=[1-2u]^\frac{-1}{2},
\end{equation}
where
\begin{equation}
1+Z_{s}=\left[1-\left(\frac{2b\sqrt{A}\pi\beta+\sqrt{\pi}\beta(\pi-2A\pi
r^{2})erf(\sqrt{A}b)+b \pi r^{2}\beta_{1}\sqrt{A}}{4\pi\sqrt{A}}\right)%
\right]^{\frac{-1}{2}}.
\end{equation}
The{\color{white}i}maximum value{\color{white}i}of the{\color{white}i}surface redshift{\color{white}i}for the compact{\color{white}i}stars is{\color{white}i}shown in
figure \textbf{9}.

\section{Discussion and conclusion}\label{Summary and conclusion}
Different{\color{white}i}cosmological observations {\color{white}i}have been{\color{white}i}shown that {\color{white}i}our Universe{\color{white}i}undergoes
two{\color{white}i}phases of{\color{white}i}accelerated expansion , early{\color{white}i}inflationary era{\color{white}i}and  late{\color{white}i}time acceleration . Dark{\color{white}i}energy in the{\color{white}i}form of exotic{\color{white}i}fluid or modified{\color{white}i}gravity or cosmological{\color{white}i}constant term{\color{white}i}are widely studied{\color{white}i}for this{\color{white}i}purpose.
Numerous{\color{white}i}forms of{\color{white}i}modified gravity{\color{white}i}proposed  and one recently{\color{white}i}proposal is to{\color{white}i}use torsion in a modified{\color{white}i}theory , called as   $f(T)$
gravity,  is{\color{white}i}one of the most{\color{white}i}popular modifications{\color{white}i}of General{\color{white}i}Relativity.

In this{\color{white}i}article, authors{\color{white}i}demonstrated analytical{\color{white}i}models of{\color{white}i}compact
stars under the framework of $%
f(T)$ gravity{\color{white}i}with the anisotropic{\color{white}i}gravitating static{\color{white}i}source. Linear{\color{white}i}form of $%
f(T)$ gravity{\color{white}i}model is{\color{white}i}considered here,{\color{white}i}Moreover the stars{\color{white}i}are
supposed to be{\color{white}i}anisotropic in{\color{white}i}their internal{\color{white}i}structure. It{\color{white}i}is
observed{\color{white}i}that the analytic{\color{white}i}solution in $f(T)$ gravity{\color{white}i}have a
matching{\color{white}i}by the interior{\color{white}i}metric with{\color{white}i}the well-known{\color{white}i}exterior metric.
 The{\color{white}i}following properties{\color{white}i}show the physical{\color{white}i}analysis of the{\color{white}i}obtained results{\color{white}i}regarding
the anisotropic{\color{white}i}compact stars in $f(T)$ gravity:

\begin{itemize}
\item EoS parameters{\color{white}i}bound are{\color{white}i}given by ${\omega}_t>0$ and ${\omega}_r>0$, that{\color{white}i}is
 steady with{\color{white}i}ordinary matter{\color{white}i}distribution in $f(T)${\color{white}i}gravity.

\item The density{\color{white}i}and pressures{\color{white}i}both are decreasing{\color{white}i}functions that{\color{white}i}attain the{\color{white}i}maximum value{\color{white}i}at the{\color{white}i}center.

\item It has{\color{white}i}been shown{\color{white}i}that the anisotropy{\color{white}i}will be directed{\color{white}i}outward when $%
p_t>p_r$ this means that $\Delta>0$ and{\color{white}i}directed inward{\color{white}i}when
$p_t<p_r$ implying $\Delta<0$. For larger{\color{white}i}values of $r$, of
different{\color{white}i}strange stars{\color{white}i}in our{\color{white}i}case $\Delta>0$, which{\color{white}i}tells that
anisotropic{\color{white}i}force is most{\color{white}i}beneficial for construction{\color{white}i}of more
massive{\color{white}i}stars in $f(T)$ gravity.

\item Variation{\color{white}i}of $\upsilon _{st}^2-\upsilon _{sr}^2$ for different{\color{white}i}strange stars{\color{white}i}shows that  $%
|\upsilon _{st}^{2}-\upsilon _{sr}^{2}|\leq1$. In this{\color{white}i}way, our
suggested strange{\color{white}i}star model{\color{white}i}shows more{\color{white}i}stablity in $f(T)${\color{white}i}gravity.
\item In this paper, we have used the diagonal tetrad to discuss the dynamics of the gravitating source. Also, one can use the off diagonal tetrad to study the compact stars.
\end{itemize}

\section{APPENDIX}\label{APPENDIX}
In this{\color{white}i}appendix we{\color{white}i}present exact{\color{white}i}forms of{\color{white}i}different quatities{\color{white}i}which have{\color{white}i}been used{\color{white}i}in our{\color{white}i}analysis.
\bigskip
\begin{eqnarray}\label{ap1}
&&\rho =\frac{b\left( 3A\sqrt{b}\left( 4+\left( 2+3br^{2}+b^{2}r^{4}\right)
\beta \right) \right) }{8\pi \left( 1+br^{2}\right) ^{3}\left( 3A\sqrt{b}%
+B\left( 2+br^{2}\right) ^{3/2}\right) } \\ \nonumber
&&+\frac{B\sqrt{2+br^{2}}\left( -2(-7+\beta )+5b^{2}r^{4}\beta
+b^{3}r^{6}\beta +2br^{2}(5+\beta )\right) }{8\pi \left( 1+br^{2}\right)
^{3}\left( 3A\sqrt{b}+B\left( 2+br^{2}\right) ^{3/2}\right) } +\frac{\beta }{8\pi r^{2}}-\frac{\beta _{1}}{16\pi },
\end{eqnarray}

\bigskip
\begin{equation}\label{ap2}
p_{r}=\frac{\left( 3A\sqrt{b}+B\sqrt{2+br^{2}}\left( 2+7br^{2}\right)
\right) \beta }{8\pi \left( r+br^{3}\right) ^{2}\left( 3A\sqrt{b}+B\left(
2+br^{2}\right) ^{3/2}\right) }-\frac{\beta }{8\pi r^{2}}-\frac{\beta _{1}}{%
16},
\end{equation}

\begin{eqnarray}\label{ap3}
p_{t} &=&\frac{b\left( -18A^{2}b\sqrt{2+br^{2}}+B^{2}\left( 2+br^{2}\right)
^{3/2}\left( 4+br^{2}+b^{2}r^{4}\right) -3A\sqrt{b}B\left( 4+br^{2}\left(
7+br^{2}\right) \right) \right) }{8\pi \left( 1+br^{2}\right) ^{3}\sqrt{%
2+br^{2}}\left( 3A\sqrt{b}+B\left( 2+br^{2}\right) ^{3/2}\right) ^{2}} \\ \nonumber
&&+\frac{3Ab\sqrt{b}B\left( 4+br^{2}\left( 7+br^{2}\right) \right) }{8\pi
\left( 1+br^{2}\right) ^{3}\sqrt{2+br^{2}}\left( 3A\sqrt{b}+B\left(
2+br^{2}\right) ^{3/2}\right) ^{2}}-\frac{{\beta }_{1}}{16\pi },
\end{eqnarray}

Also, the{\color{white}i}equation{\color{white}i}of state (EOS) parameters{\color{white}i}can be written{\color{white}i}as

\begin{equation}\label{ap4}
{\large \omega }_{r}{\large (r)}{\large =-}\frac{\left( 1+br^{2}\right) 3A%
\sqrt{b}\left( 2+br^{2}\right) \beta }{3A\sqrt{b}\left( 4+\left(
2+3br^{2}+b^{2}r^{4}\right) \beta \right) +B\sqrt{2+br^{2}}\left( 14-2\beta
+5b^{2}r^{4}\beta +b^{3}r^{6}\beta +2br^{2}(5+\beta )\right) }
\end{equation}

\begin{equation}\label{ap5}\nonumber
{\large -}\frac{B\left( 1+br^{2}\right) \sqrt{2+br^{2}}\left(
-2+4br^{2}+b^{2}r^{4}\right) \beta }{3A\sqrt{b}\left( 4+\left(
2+3br^{2}+b^{2}r^{4}\right) \beta \right) +B\sqrt{2+br^{2}}\left( 14-2\beta
+5b^{2}r^{4}\beta +b^{3}r^{6}\beta +2br^{2}(5+\beta )\right) }
\end{equation}%
\begin{eqnarray}\label{ap6}
{\large \omega }_{t}{\large (r)}{\large =} &&\frac{-18A^{2}b\sqrt{2+br^{2}}%
+B^{2}\left( 2+br^{2}\right) ^{3/2}\left( 4+br^{2}+b^{2}r^{4}\right) }{\sqrt{%
2+br^{2}}\left( 3A\sqrt{b}+B\left( 2+br^{2}\right) ^{3/2}\right) } \\ \nonumber
&&\times \frac{1}{\left( 3A\sqrt{b}\left( 4+\left( 1+br^{2}\right) \left(
2+br^{2}\right) \beta \right) +B\sqrt{2+br^{2}}\left( 14-2\beta
+br^{2}\left( 10+\left( 2+br^{2}\left( 5+br^{2}\right) \right) \beta \right)
\right) \right) }
\end{eqnarray}

\begin{eqnarray}\label{ap7}\nonumber
&&-\frac{1}{\sqrt{2+br^{2}}\left( 3A\sqrt{b}+B\left( 2+br^{2}\right)
^{3/2}\right) } \\ \nonumber
\times &&\frac{3A\sqrt{b}B\left( 4+br^{2}\left( 7+br^{2}\right) \right) }{%
\left( 3A\sqrt{b}\left( 4+\left( 1+br^{2}\right) \left( 2+br^{2}\right)
\beta \right) +B\sqrt{2+br^{2}}\left( 14-2\beta +br^{2}\left( 10+\left(
2+br^{2}\left( 5+br^{2}\right) \right) \beta \right) \right) \right) }
\end{eqnarray}

\begin{equation*}\label{ap8}
\frac{d^{2}\rho }{dr^{2}}=\frac{3b^{2}(27A^{3}b^{3/2}\left( 2+br^{2}\right)
^{3/2}\left( -4-\beta +br^{2}\left( 28+\left( 3+br^{2}\left( 5+br^{2}\right)
\right) \beta \right) \right) }{\left( 4\pi \left( 1+br^{2}\right)
^{5}\left( 2+br^{2}\right) ^{3/2}\left( 3A\sqrt{b}+B\left( 2+br^{2}\right)
^{3/2}\right) ^{3}\right) }
\end{equation*}

\begin{equation*}
+\frac{B^{3}3b^{2}\left( 2+br^{2}\right) ^{3}\left( -52+12\beta
+br^{2}\left( 250-62\beta +br^{2}\left( 538-156\beta +br^{2}\left(
338-89\beta \right) \right) \right) \right) }{\left( 4\pi \left(
1+br^{2}\right) ^{5}\left( 2+br^{2}\right) ^{3/2}\left( 3A\sqrt{b}+B\left(
2+br^{2}\right) ^{3/2}\right) ^{3}\right) }
\end{equation*}

\begin{equation*}
+\frac{B^{3}3b^{2}\left( 2+br^{2}\right) ^{3}br^{2}\left( 70+\left(
3+br^{2}\left( 11+br^{2}\right) \right) \beta \right) }{\left( 4\pi \left(
1+br^{2}\right) ^{5}\left( 2+br^{2}\right) ^{3/2}\left( 3A\sqrt{b}+B\left(
2+br^{2}\right) ^{3/2}\right) ^{3}\right) }
\end{equation*}

\begin{equation*}
+\frac{9b^{2}A\sqrt{b}B^{2}\left( 2+br^{2}\right) ^{3/2}\left( 10(-13+\beta
)+br^{2}\left( 622-58\beta +br^{2}\left( 1214-68\beta +3br^{2}\left(
226+23\beta \right) \right) \right) \right) }{\left( 4\pi \left(
1+br^{2}\right) ^{5}\left( 2+br^{2}\right) ^{3/2}\left( 3A\sqrt{b}+B\left(
2+br^{2}\right) ^{3/2}\right) ^{3}\right) }
\end{equation*}

\begin{equation*}
+\frac{9b^{2}A\sqrt{b}B^{2}\left( 2+br^{2}\right) ^{3/2}br^{2}\left(
40+\left( 33+br^{2}\left( 11+br^{2}\right) \right) \beta \right) }{\left(
4\pi \left( 1+br^{2}\right) ^{5}\left( 2+br^{2}\right) ^{3/2}\left( 3A\sqrt{b%
}+B\left( 2+br^{2}\right) ^{3/2}\right) ^{3}\right) }
\end{equation*}

\begin{equation*}
+\frac{27A^{2}Bb^{3}\left( -10(11+\beta )+br^{2}\left( 566-2\beta
+br^{2}\left( 2(521+52\beta )+3br^{2}\left( 186+63\beta \right) \right)
\right) \right) }{\left( 4\pi \left( 1+br^{2}\right) ^{5}\left(
2+br^{2}\right) ^{3/2}\left( 3A\sqrt{b}+B\left( 2+br^{2}\right)
^{3/2}\right) ^{3}\right) }
\end{equation*}

\begin{equation*}
+\frac{27A^{2}Bb^{4}r^{2}\left( 32+\left( 41+br^{2}\left( 11+br^{2}\right)
\right) \beta \right) }{\left( 4\pi \left( 1+br^{2}\right) ^{5}\left(
2+br^{2}\right) ^{3/2}\left( 3A\sqrt{b}+B\left( 2+br^{2}\right)
^{3/2}\right) ^{3}\right) }
\end{equation*}

\begin{eqnarray}\label{ap9}
\frac{d^{2}p_{r}}{dr^{2}} &=&-\frac{3b^{2}\left( 27A^{3}b^{3/2}\left(
2+br^{2}\right) ^{3/2}\left( -1+br^{2}\left( 4+br^{2}\right) \right) \beta
\right) }{\left( 4\pi \left( 1+br^{2}\right) ^{4}\left( 2+br^{2}\right)
^{3/2}\left( 3A\sqrt{b}+B\left( 2+br^{2}\right) ^{3/2}\right) ^{3}\right) }
\notag \\
&&-\frac{3b^{2}\left( 3A\sqrt{b}B^{2}\left( 2+br^{2}\right) ^{3/2}\left(
10+br^{2}\left( -68+3b^{2}r^{4}\left( 23+br^{2}\left( 10+br^{2}\right)
\right) \right) \right) \beta \right) }{\left( 4\pi \left( 1+br^{2}\right)
^{4}\left( 2+br^{2}\right) ^{3/2}\left( 3A\sqrt{b}+B\left( 2+br^{2}\right)
^{3/2}\right) ^{3}\right) }  \notag \\
&&-\frac{3b^{2}\left( B^{3}\left( 2+br^{2}\right) ^{3}\left( 12+br^{2}\left(
-74+br^{2}\left( -82+br^{2}\left( -7+br^{2}\left( 10+br^{2}\right) \right)
\right) \right) \right) \beta \right) }{\left( 4\pi \left( 1+br^{2}\right)
^{4}\left( 2+br^{2}\right) ^{3/2}\left( 3A\sqrt{b}+B\left( 2+br^{2}\right)
^{3/2}\right) ^{3}\right) }  \notag \\
&&-\frac{3b^{2}\left( 9A^{2}bB\left( -10+br^{2}\left( 8+3br^{2}\left(
32+br^{2}\left( 31+br^{2}\left( 10+br^{2}\right) \right) \right) \right)
\right) \beta \right) }{\left( 4\pi \left( 1+br^{2}\right) ^{4}\left(
2+br^{2}\right) ^{3/2}\left( 3A\sqrt{b}+B\left( 2+br^{2}\right)
^{3/2}\right) ^{3}\right) }
\end{eqnarray}%

\begin{eqnarray}\label{ap10}
&&\Delta =\frac{b\left( -18A^{2}b\sqrt{2+br^{2}}+B^{2}\left( 2+br^{2}\right)
^{3/2}\left( 4+br^{2}+b^{2}r^{4}\right) -3A\sqrt{b}B\left( 4+br^{2}\left(
7+br^{2}\right) \right) \right) }{8\pi \left( 1+br^{2}\right) ^{3}\sqrt{%
2+br^{2}}\left( 3A\sqrt{b}+B\left( 2+br^{2}\right) ^{3/2}\right) ^{2}} \\ \nonumber
&&-\frac{\left( 3A\sqrt{b}+B\sqrt{2+br^{2}}\left( 2+7br^{2}\right) \right)
\beta }{8\pi \left( r+br^{3}\right) ^{2}\left( 3A\sqrt{b}+B\left(
2+br^{2}\right) ^{3/2}\right) }
\end{eqnarray}%

\begin{equation}
\frac{d\rho }{dr}=-\frac{b^{2}r\left( 9A^{2}b\sqrt{2+br^{2}}\left( 12+\left(
1+br^{2}\right) \left( 3+br^{2}\right) \beta \right) \right) }{\left( 4\pi
\left( 1+br^{2}\right) ^{4}\sqrt{2+br^{2}}\left( 3A\sqrt{b}+B\left(
2+br^{2}\right) ^{3/2}\right) ^{2}\right) }
\end{equation}

\begin{equation}\nonumber
-\frac{b^{2}r\left( B^{2}\left( 2+br^{2}\right) ^{3/2}\left( 78-18\beta
+br^{2}\left( 96-20\beta +br^{2}\left( 30+\left( 5+br^{2}\left(
8+br^{2}\right) \right) \beta \right) \right) \right) \right) }{\left( 4\pi
\left( 1+br^{2}\right) ^{4}\sqrt{2+br^{2}}\left( 3A\sqrt{b}+B\left(
2+br^{2}\right) ^{3/2}\right) ^{2}\right) }
\end{equation}

\begin{equation}\nonumber
-\frac{3A\sqrt{b}B\left( 3(39+\beta )+br^{2}\left( 126+26\beta +br^{2}\left(
33+\left( 37+2br^{2}\left( 8+br^{2}\right) \right) \beta \right) \right)
\right) }{\left( 4\pi \left( 1+br^{2}\right) ^{4}\sqrt{2+br^{2}}\left( 3A%
\sqrt{b}+B\left( 2+br^{2}\right) ^{3/2}\right) ^{2}\right) }
\end{equation}

\begin{equation}
\frac{dp_{r}}{dr}=\frac{b^{2}r\left( 9A^{2}b\sqrt{2+br^{2}}\left(
3+br^{2}\right) +B^{2}\left( 2+br^{2}\right) ^{3/2}\left( -18+br^{2}\left(
-2+br^{2}\left( 7+br^{2}\right) \right) \right) \right) \beta }{4\pi \left(
1+br^{2}\right) ^{3}\sqrt{2+br^{2}}\left( 3A\sqrt{b}+B\left( 2+br^{2}\right)
^{3/2}\right) ^{2}}
\end{equation}

\begin{equation*}
+\frac{3A\sqrt{b}B\left( 3+br^{2}\left( 23+2br^{2}\left( 7+br^{2}\right)
\right) \right) \beta }{4\pi \left( 1+br^{2}\right) ^{3}\sqrt{2+br^{2}}%
\left( 3A\sqrt{b}+B\left( 2+br^{2}\right) ^{3/2}\right) ^{2}}
\end{equation*}



\end{document}